\documentclass[preprint,12pt]{elsarticle}

\usepackage{epsfig}

\usepackage{amssymb}

\journal{ }

\begin{document}

\begin{frontmatter}
\title{Key Modes for Time-Space Evolutions of ENSO and PDO by ESMD Method\footnote{I note that this work had been finished on December, 09, 2016, and the patent application for ``\emph{Integrated ESMD Method for space-time data analysis}" had been submitted to the patent office of the People's Republic of China on January, 11, 2019.
   \textbf{Relative to the conventional empirical orthogonal function (EOF) approach, the advantage of the integrated version of our ESMD method goes without saying (please see the fresh figures in the last)!}  }}


\author{Jin-Liang Wang}

\address{ {Research Institute for ESMD method and Its Applications,}\\
{College of science, Qingdao University of Technology,
Shandong, China (266520).}\\
 {E-mail: wangjinliang0811@126.com}
}
\begin{abstract}
 The relation between El Ni\~{n}o-Southern Oscillation (ENSO) and Pacific Decadal Oscillation (PDO) perplexes the researchers. Probably, this is due to the default choice on empirical orthogonal function (EOF) method whose outputs are merely static maps without evolution characteristics. To change this situation, the new extreme-point symmetric mode decomposition (ESMD) method is tried. It reveals that: (1) As the ENSO concerned, the key modes are the quarter-decadal and half-decadal ones; (2) As the PDO concerned, the key modes are the half-decadal, decadal and interdecadal ones; (3) The half-decadal mode can be seen as a mutual component for ENSO and PDO and it plays a key part in the so-called ``ENSO-like" variation of PDO. In addition, an integrated version of ESMD is developed for time-space evolution. Its outputs have verified the warm-cold shiftings of ENSO and PDO for the most typical episode during 1997-2000.
  \end{abstract}

\end{frontmatter}



\baselineskip = 18 pt

\section{Introduction}
\setcounter{equation}{0}

The ENSO, varying between anomalous warm (El Ni\~{n}o) and cold (La Ni\~{n}a) states in the tropical Pacific, is a prominent climate phenomenon affecting extreme weather conditions worldwide (Cai \textit{et al} 2015,  Yeh \textit{et al} 2009, Enfield 1989).
  The PDO, shifting between anomalous warm and cold regimes in the North Pacific,
  is also a prominent one (Mantua \textit{et al} 1997, Nathan and Steven 2002). Together with ENSO, the PDO becomes an important target of ongoing
 research within the meteorological and climate dynamics communities, and they are central to
 the work of many geologists, ecologists, natural resource managers, and social scientists (Newman \textit{et al} 2016).

There is a popular belief that the ENSO and PDO differ in their time scales. The ENSO cycle has a significant interannual variation
with period of about $2$-$8$ yr [$2$-$7$ yr (Capotondi \textit{et al} 2015, Collins \textit{et al} 2010), $3$-$7$ yr (White and Cayan 1998) or $3$-$8$ yr (Deser \textit{et al} 2012)].
Yet that for PDO has a significant decadal or interdecadal variation
with period of about $9$-$30$ yr [$15$-$25$ yr (Minobe 1999),
$15$-$20$ yr (Chao \textit{et al} 2000), $9$-$25$ yr (Tourre \textit{et al} 2001), $10$-$17$ yr (Lluch-Cota \textit{et al} 2003) or $10$-$30$ yr (Yang and Zhang 2003)].
We note that, in addition to the $15$-$25$ yr band, Minobe (1999) had also found another one with $50$-$70$ yr
which accords with the documented cycle with duration $52$ yr delimited by
the typical regime-shift years $1925$, $1947$ and $1977$ (Mantua \textit{et al} 1997).
But, as argued by Chao \textit{et al} (2000), the regime shifts also occur around $1941$-$42$ and $1957$-$58$
and the significant variation of PDO should be in a shorter time scale.
Notice that the PDO represent not a single physical mode but rather a combination of many processes (Newman \textit{et al} 2016, Newman 2007, Chen and Wallace 2016), the ENSO
may be also a combined one. A more convincing viewpoint is:
 ``\textit{Both the occurrences of PDO and ENSO are due to the superposition of many intrinsic modes with different time scales}".
Certainly, how to find these modes is a challenge to the researchers and it becomes the topic of the present article.

It is also known that the PDO is usually described as a long-lived ``ENSO-like" variation
 due to the similarity of its typical spatial-pattern to that of ENSO (Newman \textit{et al} 2016, Zhang \textit{et al} 1997, Chen and Wallace 2015).
 This similarity perplexes the researchers.
  One paradigm views the PDO as an independent phenomenon centered
in the North Pacific, while another regards it as a largely reddened response to ENSO forcing from the tropics (Shakun and Shaman 2009).
Comparatively, the latter viewpoint attracts more supporters (Shakun and Shaman 2009, Newman \textit{et al} 2003, Deser \textit{et al} 2004, Lorenzo \textit{et al} 2010, Vimont 2005), after all the pattern similarity implies a potential correlation.
This debate has been continued  and the question ``\textit{does the PDO exist independent of the ENSO cycle?}"
is still in suspense till now (Chen and Wallace 2015). Probably, the PDO and ENSO are two aspects of one thing
and their relation can not be understood as one dominating another.
From the viewpoint of mode decomposition, there may exist mutual interannual modes for them
 which result in pattern similarity and, beside of the mutual ones, the PDO may also possess other specific modes
 which contribute to decadal or interdecadal oscillations.

 There is an essential difficulty for describing ENSO and PDO that the sea surface temperature (SST) varies not only with time but also with space.
 For the sake of statistical forecasting, Lorenz (1956) proposed the EOF method which
 has now been borrowed as the default approach for time-space data analyses.
 It can output several spatial modes together with same amount of principal components (PCs).
 By the way, according to the decomposition rule, the PCs only reflect the intensity-changes of their corresponding spatial modes
 and they can not be seen as intrinsic temporal modes of the original time series.
  For example, in narrow sense the PDO is the leading PC of EOF respect to the SST anomalies (Mantua \textit{et al} 1997). Its corresponding spatial mode is a static map with low temperature in the central North Pacific and high temperature along the west coast of North America (or the inverse). In the same way, the EOF method also yields a static map to the ENSO relative to certain historical data.
    Notice that these static maps have no evolution characteristics, it is nothing strange for the
  existence of difficulty in distinguishing PDO and ENSO. Moreover, as reviewed by Cai \textit{et al} (2015), debates also persist as to whether the central Pacific El Ni\~{n}o, whose spatial pattern resembles the second spatial mode of EOF, is part of the ENSO asymmetry or a distinct mode. In fact, Monahan \textit{et al} (2009) had already pointed out the shortcomings
  of EOF method: ``\textit{its individual spatial mode neither corresponds to individual dynamical mode nor be statistically independent of other ones}". In addition, the EOF method can not be used solely.
  It needs not only a ``detrending" technique to eliminate the non-stationarity (such as the global-warming effect) but also a ``running-mean" processing to filter out the high-frequency signals (such as the seasonal variation).
    In view of the fact that ENSO cycle varies between El Ni\~{n}o and La Ni\~{n}a states in interannual time scales, its spatial pattern should adjust accordingly. Similarly, that for PDO should also adjust in a slower manner.
  Hence, to reflect their time-space evolutions objectively it requires a method innovation.

 In the recent years we have developed the ESMD method for data analysis (Wang and Li 2013, 2015), which has advantage in revealing the multi-scale variations for non-stationary time series.
 In the present article we use it to explore the key modes for ENSO and PDO.
 To reflect the spatial variations, an integrated version of ESMD is also developed here.
 Its outputs differ from the static maps given by EOF method and the spatial patterns
 actually possess real-time adjusting characteristics. These are the real time-space evolutions.

\section{Data and Methods}
The observed monthly SST data during 1880-2015 are from the improved Extended Reconstructed Sea Surface Temperature version 4 (ERSSTv4,
http://www.ncdc.noaa.gov/data-access/marineocean-data) from the National Climate Data Center of the US.
The Ni\~{n}o-3.4 index is from the Climate Prediction Center [monthly ERSSTv4 (1981-2010 base period) Ni\~{n}o 3.4,
http://www.cpc.noaa.gov/data/indices].

The ESMD algorithms are taken from our previous works (Wang and Li 2013, 2015). By the way, the simple ESMD guideline and its corresponding software are provided in a free-access manner with a link (http://blog.sciencenet.cn/blog-686810-691899.html).
The integrated version of ESMD is implemented as follows:
(1) For each grid point, to execute the ESMD decomposition on its corresponding time series and get a series of temporal modes;
(2) To unify the number of modes, for example, to limit it to be $5$. If the number is more than $5$ then take the sum of the fifth and the latter ones as Mode 5, for the inverse case assign $0$ to the missing ones in the direct way;
(3) To fix the time, collect the corresponding values of the interested modes for all the grid points
and plot them on the whole region. For example, respect to Jan.1997 one can pick the corresponding value from the sum of Mode 2-5
to each grid point and integrate them into the first subgraph of \textsf{Fig.4}; (4) To change the time, repeat the above process and get a series of spatial patterns.

\section{Main Results}
To analyze the temporal variations of ENSO and PDO, the grid point A ($0^{\circ}$, $130^{\circ}$W) on the equator and
the grid point B ($42^{\circ}$N, $130^{\circ}$W) in the North Pacific are taken as representatives (denoted by two circles in \textsf{Fig.5}).
By executing the ESMD method on the monthly-mean SST time series (from Jan.1880 to Dec.2015) for point A it yields some decomposition results in \textsf{Fig.1}. Similar results can be gotten for point B (includes $6$ modes and a remainder R, the figure is omitted here).
Here the first modes accord with the strongest seasonal signals and their mean variation-ranges
(from winter to summer) attain $2.5^{\circ}$C and $7.0^{\circ}$C
at points A and B separately. This indicates that the seasonal change in the north extretropic is more obvious than that on the equator.
It follows from the last remainder Rs that the global-warming effects in the tropic and extretropic regions are very remarkable.
Particularly, in the latest $50$ years their total increments attain $0.65^{\circ}$C and $0.58^{\circ}$C separately, which are in line with the documented results (Hansen \textit{et al} 2006, 2010).

The feasibility of decomposition is checked in \textsf{Fig.2}.
It follows from \textsf{Fig.2a} that the sum of Mode 2-5 for point A accords well with the known Ni\~{n}o-3.4 index
(their correlation coefficient attains $0.84$). This indicates the decomposition by ESMD method is feasible and, except the seasonal and global-warming signals, the left modes with different time scales all contribute to the ENSO. The curves in \textsf{Fig.2b\&c} display the sum of Mode 3-6 for B and the sum of Mode 3-5 for A separately.
The first one possesses many zero-points which are the candidates for regime-shift times.
Particularly, three of them accord with the typical regime-shift years 1925, 1947 and 1977
(the results given here are only about single point and the more objective shifting times should be
judged by the whole spatial-evolution figures).
In addition, there is also a zero-point at 1999 which accords with the shift year given by Henley \textit{et al} (2015).
We note that, if Mode 3 is abandoned then the zero-points will differ much from these typical years (the figure is omitted).
These consistencies imply that the decomposed Mode 3-6 all contribute to the PDO.
Moreover, it follows from \textsf{Fig.2c} that, coincidentally,
these typical years also almost correspond to regime-shift thresholds for point B.
This implies the existence of synchronous change for the tropic and extretropic regions
 (As the grid points A and B concerned, the total correlation coefficient for them is $0.35$).

To apply the ``direct interpolating (DI)" approach [the second part of ESMD method] on the decomposed results it yields
the time-period distributions for the modes in \textsf{Fig.3}.
Together with the decompositions for A and B, this figure further derives a statistic result in \textsf{Tabel 1}.
We note that, differing from the conventional power-spectral methods, the outputted period for the mode here is not a fixed value anymore,
it adjusts with time in an instantaneous manner. So the listed period-ranges here are the exact variation ones, not
the estimation-intervals with errors.

It follows from \textsf{Tabel 1} that, except Mode 1 (the seasonal signal),
Mode 2 \& 3 are the first and second strongest interannual oscillating-components.
Notice that their mean periods to A and B are very closer ($2.6$ yr vs $2.4$ yr and $5.5$ yr vs $5.1$ yr),
we take the terms ``quarter-decadal" and ``half-decadal" as time-scales to differ them.
Mode 4 can be seen as a decadal component whose correlation-coefficient
between A and B is higher than the previous two.
As for Mode 5, though its time-scale has some difference on these two points, it can be seen as an interdecadal variation with rough range
$18$-$40$ yr (only in the time-segment 1880-1895, that for B is lower than $18$ yr).
A relatively strong correlation between the equator and North Pacific is also implied by this mode.
The last mode is another interdecadal one associated with point B solely. Its period ranges from $42$ to $72$ yr (with a mean $50$ yr) which approximates the one in $50$-$70$ yr given by Minobe (1999). Yet, since the amplitude of it is the smallest one among all the modes, its contribution to PDO should be the least.

Among all the modes, the half-decadal one is most attractive. It may be the key component
connecting the ENSO and PDO.
To retrieve \textsf{Fig.2b} we see this mode has made great contribution to regime shift.
 In fact, the so-called ``decadal oscillation" does not exclude the interannual variations.
To trace the source of the highly cited term ``ENSO-like" for PDO,
it was firstly proposed to indicate the low-pass oscillating component of leading PC
which is filtered with a threshold $6$ yr (Zhang \textit{et al} 1997). So it is rational to take the half-decadal mode
as a component of PDO.

The duration of 1997-2000 is the most typical episode for ENSO and PDO since,
during this time, there is not only a strong shift from El Ni\~{n}o to La Ni\~{n}a
in the tropic Pacific but also a strong shift from warm to cold regimes in the North Pacific.
By the way, the recent episode 2013-2015 is not a typical one (the ENSO cycle is half-baked), though
there exists a strong regime-shift (see \textsf{Fig.2}).
With respect to this typical episode, the application of integrated ESMD method
yields two series of spatial patterns for them.
The \textsf{Fig.4} exhibits a time-space evolution of ENSO on the whole Pacific, where the conflict
between anomalous warm and cold signals is reflected in a distinct way.
The \textsf{Fig.5} exhibits the similar result for PDO, where only Mode 3
is used. It confirms the saying that the half-decadal mode is a component of PDO.
But there is more to it than this. In fact, as the case concerned, the so-called ``regime shift" of PDO is completely
dominated by this mode.

\section{Conclusions}

The application of ESMD method on monthly-mean SST time series reveals two series of temporal modes
for one representative grid point on the equator and another in the North Pacific.
Besides the remainder modes which reflect the global-warming effects, they all possess
the modes in seasonal, quarter-decadal, half-decadal, decadal and interdecadal time scales.

As the ENSO concerned, except the seasonal and global-warming signals,
all the left modes make contribution to it. Its key modes are the quarter-decadal and half-decadal ones.
Relatively, the contributions from decadal and interdecadal modes are very small.
This is why the ENSO is commonly seen as an event with interannual time scale.

As the PDO concerned, the quarter-decadal signal is abandoned as well as the seasonal and global-warming ones.
The time-space evolution figures confirm that the half-decadal mode is a key component to it.
In addition to this, the decadal and interdecadal modes also have considerable contributions.
This is why the PDO is commonly seen as an event with decadal or interdecadal time scale.

In addition, the half-decadal mode can be seen as a mutual oscillating-component
for ENSO and PDO and it plays a key part in the so-called ``ENSO-like" variation of PDO.

By the way, enlightened by the shifting between anomalous warm and cold signals in North Pacific
(see \textsf{Fig.4\& 5}), we give two conjectures here:\\
(1)  The North Pacific Current (flows from west to east between $35$-$42^{\circ}$N) is probably dominated by
the warm Kuroshio Current (from the south) and cold Oyashio Current (from the north) in turn with
a cycle of about $5$ years.
In case the warm water attains the west coast of North America (followed by cold water near the central North Pacific)
the warm regime of PDO occurs. For the inverse case, it is the cold regime.\\
(2) In case the southward California Current (at the east boundary) is full of warm water from Kuroshio Current, it may
create an advantageous condition for the occurrence of El Ni\~{n}o (perhaps by impacting the tropic currents or
by compelling the atmospheric circulation).
[On November 24, 2019, I note that in another article finished on November 20, 2017 with title ``Current-Generating Mechanism for El Ni\~{n}o and La Ni\~{n}a, an Data Evidence Given by Integrated ESMD Method", there is a detailed study. The warm water from the high-temperature sea-area off Central America plays a key part.]

\vskip 3mm
 \bibliographystyle{ametsoc2014}
 \bibliography{references}

 Cai, W. J., and Coauthors, 2015: ENSO and greenhouse warming. \textit{Nature Clim. Change}, \textbf{5}, 849-859, doi: 10.1038 /NCLIMATE2743.

 Capotondi, A. and coauthors, 2015: Understanding ENSO diversity. \textit{Bull. Amer. Meteor. Soc.}, \textbf{96}, 921-938,
doi: 10.1175 /BAMS-D- 13-00117.1.

 Chao, Y., M. Ghil, and J. C. McWilliams, 2000: Pacific interdecadal variability in this centurys sea surface temperatures.
\textit{Geophys. Res. Lett.}, \textbf{27}(15), 2261-2264, doi: 10.1029 /1999GL011324.

 Chen, X. Y., and J. M. Wallace, 2015: ENSO-like variability: 1900-2013. \textit{J. Climate}, \textbf{28}, 9623-9641, doi:10.1175 /JCLI-D-15-0322.1.

 Chen, X. Y., and J. M. Wallace, 2016: Orthogonal PDO and ENSO indices. \textit{J. Climate}, \textbf{29}, 3883-3892,
doi: 10.1175 /JCLI-D-15-0684.1.

 Collins, M., and Coauthors, 2010: The impact of global warming on the tropical Pacific ocean and El Ni\~{n}o. \textit{Nature Geosci.}, \textbf{3}, 391-397, doi:10.1038 /NGEO868.

 Deser, C., and Coauthors, 2012: ENSO and Pacific decadal variability in the
community climate system model version 4. \textit{J. Climate}, \textbf{25}, 2622-2651,
doi:10.1175 /JCLI-D-11-00301.1.

 Deser, C., A. S. Phillips, and J. W. Hurrell, 2004: Pacific interdecadal climate
variability: Linkages between the tropics and the North Pacific during boreal winter
since 1900. \textit{J. Climate}, \textbf{17}, 3109-3124.

 Enfield, D. B., 1989: El Ni\~{n}o, past and present. \textit{Reviews of Geophysics}, \textbf{27}(1) 159-187.

 Hansen, J., R. Ruedy, M. Sato, and K. Lo, 2010: Global surface temperature change. \textit{Reviews of Geophysics},
 \textbf{48}, RG4004, doi: 10.1029 /2010RG000345.

 Hansen, J., M. Sato, R. Ruedy, K. Lo, D. W. Lea, and M. Medina-Elizade, 2006: Global temperature change. \textit{PNAS}, \textbf{103} (39), 14288-14293.

 Henley, B. J., J. Gergis, D. J. Karoly, S. B. Power, J. Kennedy, and C. K. Folland, 2015: A
tripole index for the interdecadal Pacific oscillation. \textit{Clim. Dyn.}, \textbf{45}(11), 3077-3090,
doi: 10.1007 /s00382-015-2525-1.

 Lluch-Cota, D. B., W. S. Wooster, S. R. Hare, D. Lluch-Belda, and A. Par\'{e}s-Sierra, 2003: Principal modes and related frequencies of sea surface temperature variability in the Pacific coast of North America. \textit{Journal of Oceanography}, \textbf{59}, 477-488.

 Lorenz, E. N., 1956: \textit{Empirical Orthogonal Functions and Statistical Weather Prediction}. Statistical Forecasting Project Report 1, MIT Department of Meteorology, Cambridge, 49 pp.

 Lorenzo E. D., K. M. Cobb, J. Furtado, N. Schneider, B. Anderson, A. Bracco, M. A.
Alexander, and D. Vimont, 2010: Central Pacific El Ni\~{n}o and decadal climate change in
 the North Pacific. \textit{Nature Geosci.}, \textbf{3}(11), 762-765, doi: 10.1038 /NGEO984.

 Mantua, N. J., S. R. Hare, Y. Zhang, J. M. Wallace, and R. C. Francis,
1997: A Pacific interdecadal climate oscillation with impacts on
salmon production. \textit{Bull. Amer. Meteor. Soc.}, \textbf{78},
1069-1079.

 Minobe, S., 1999: Resonance in bidecadal and pentadecadal climate oscillations over the North Pacific:
 Role in climatic regime shifts. \textit{Geophys. Res. Lett.}, \textbf{26}, 855-858.

 Monahan, A. H., J. C. Fyfe, M. H. P. Ambaum, D. B. Stephenson, and G. R. North, 2009:
 Empirical orthogonal functions: the medium is the message. \textit{J. Climate}, \textbf{22}, 6501-6514, doi: 10.1175 /2009JCLI3062.1.

 Nathan, J. M. and R. H. Steven, 2002: The Pacific decadal oscillation. \textit{Journal of Oceanography}, \textbf{58}, 35-44.

 Newman, M., G. P. Compo, and M. Alexander, 2003: ENSO-forced variability of the
Pacific decadal oscillation. \textit{J. Climate}, \textbf{16}, 3853-3857.

 Newman, M., 2007: Interannual to decadal predictability of
tropical and North Pacific sea surface temperatures. \textit{J. Climate}, \textbf{20},
2333-2356, doi:10.1175/JCLI4165.1.

 Newman, M., and coauthors, 2016: The Pacific decadal oscillation, revisited. \textit{J. Climate}, \textbf{29}, 4399-4427,
doi:10.1175/JCLI-D-15-0508.1.

 Shakun, J. D., and J. Shaman, 2009: Tropical origins of North and South Pacific decadal
variability. \textit{Geophys. Res. Lett.}, \textbf{36}, L19711, doi:10.1029 /2009 GL040313.

 Tourre, Y. M., B. Rajagopalan, Y. Kushnir, M. Barlow, and W. B. White, 2001:
 Patterns of coherent decadal and interdecadal climate signals in the Pacific basin during the
20th century. \textit{Geophys. Res. Lett.}, \textbf{28}, 2069-2072.

 Vimont, D. J., 2005: The contribution of the interannual ENSO cycle to the spatial
 pattern of decadal ENSO-like variability. \textit{J. Climate}, \textbf{18}, 2080-2092.

 Wang, J. L., and Z. J. Li, 2013:  Extreme-point symmetric mode decomposition method for data analysis. \textit{Advances in Adaptive Data Analysis}, \textbf{5}(3), 1350015, doi: 10.1142/S1793536913500155.

 Wang, J. L., and Z. J. Li, 2015: \textit{Extreme-Point Symmetric Mode Decomposition Method: A New Approach for Data Analysis and Science Exploration}. Beijing: Higher Education Press (in Chinese).

 White, W. B. and D. R. Cayan, 1998: Quasi-periodicity and global symmetries in interdecadal upper ocean temperature variability.
 \textit{J. Geophys. Res.}, \textbf{103}(C10), 21335-21354.

 Yang, H. J. and Q. Zhang, 2003: On the decadal and interdecadal variability in the Pacific ocean. \textit{Advances in Atmospheric Science}. \textbf{20} (2), 173-184.

 Yeh, S. W., J. S. Kug, B. Dewitte, M. H. Kwon, B. P. Kirtman, and F. F. Jin, 2009: El Ni\~{n}o in a changing climate. \textit{Nature}, \textbf{461}, 511-514, doi: 10.1038 /nature08316.

 Zhang, Y., J. M. Wallace, and D. S. Battisti, 1997: ENSO-like
interdecadal variability: 1900-93. \textit{J. Climate}, \textbf{10}, 1004-1020.

\newpage
\begin{table}
\caption{The averaged periods ($P_{A}$ \& $P_{B}$, with unit ``yr") and amplitudes ($A_{A}$ \& $A_{B}$,
 with unit ``$^{\circ}$C") for the modes to points A and B respectively. The corresponding ranges for them are denoted by $RP_{A}$, $RP_{B}$, $RA_{A}$ and $RA_{B}$.
 The items in the last column ($C_{AB}$) are the correlation coefficients for the modes to A and B.}
 \vskip 2mm
{\small \begin{tabular*}{\hsize}{@{\extracolsep\fill}lcccccccccc@{}}
\    Modes &$P_{A}$ &$RP_{A}$ &$P_{B}$ &$RP_{B}$ &$A_{A}$ &$RA_{A}$ &$A_{B}$ &$RA_{B}$ &$C_{AB}$ \\
\    Mode1 &0.7 &0.2--1.3 &0.9 &0.2--1.2 &1.26 &0.02--2.30 &3.50 &2.56--4.30 &-0.57 \\
\    Mode2 &2.6 &0.8--5.6 &2.4 &0.9--4.5 &0.83 &0.00--2.06 &0.45 &0.01--1.62 &0.32 \\
\    Mode3 &5.5 &2.6--8.9 &5.1 &2.6--8.1 &0.56 &0.02--1.43 &0.33 &0.03--1.29 &0.35 \\
\    Mode4 &13  &8--17    &11  &6--19    &0.32 &0.16--0.59 &0.27 &0.02--0.57 &0.46 \\
\    Mode5 &30  &24--39   &22  &13--33   &0.25 &0.16--0.43 &0.29 &0.16--0.73 &0.54 \\
\    Mode6 &--  &--       &50  &42--72   &--   &--         &0.20 &0.03--0.45 &-- \\
\end{tabular*}}
\end{table}


\begin{figure}[h]
\centerline{\includegraphics[width=\textwidth]{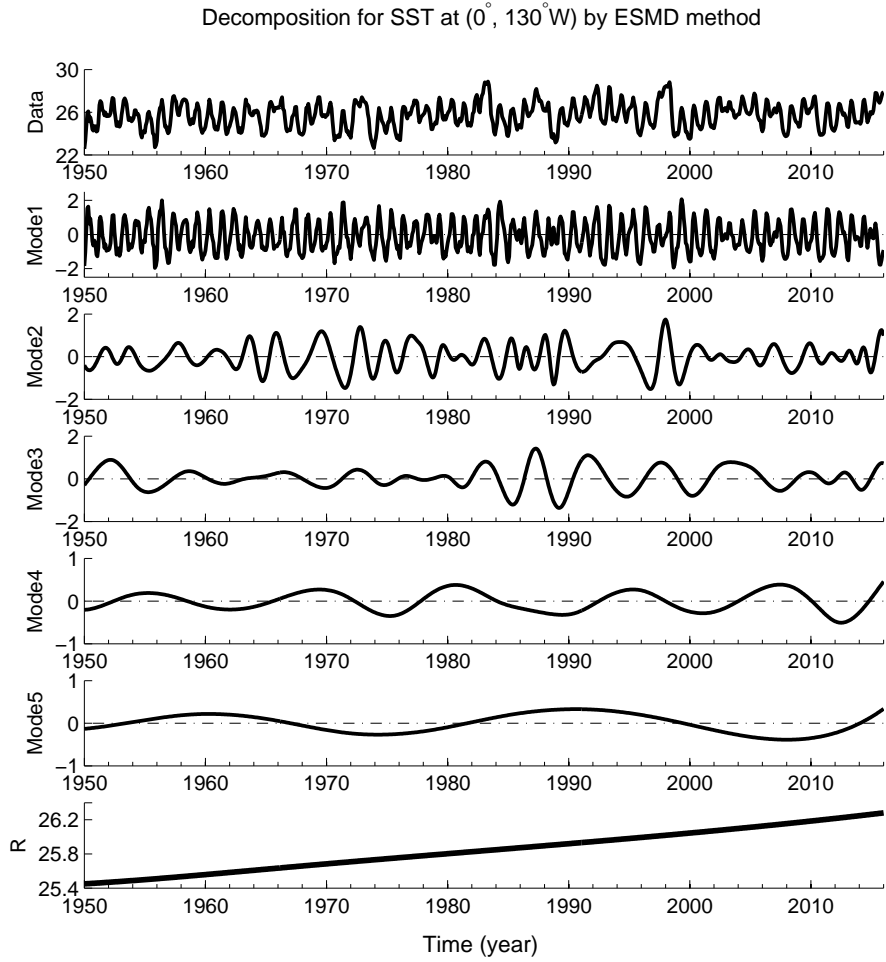}}

\caption{Part of the decomposition result for the SST to grid point
A ($0^{\circ}$, $130^{\circ}$W) by ESMD method. ``Data" denotes the raw SST
data, Mode 1-5 are the decomposed modes. Here the vertical axes
possess the same unit $^{\circ}$C.}
\end{figure}

\begin{figure}[h]
\centerline{\includegraphics[width=\textwidth]{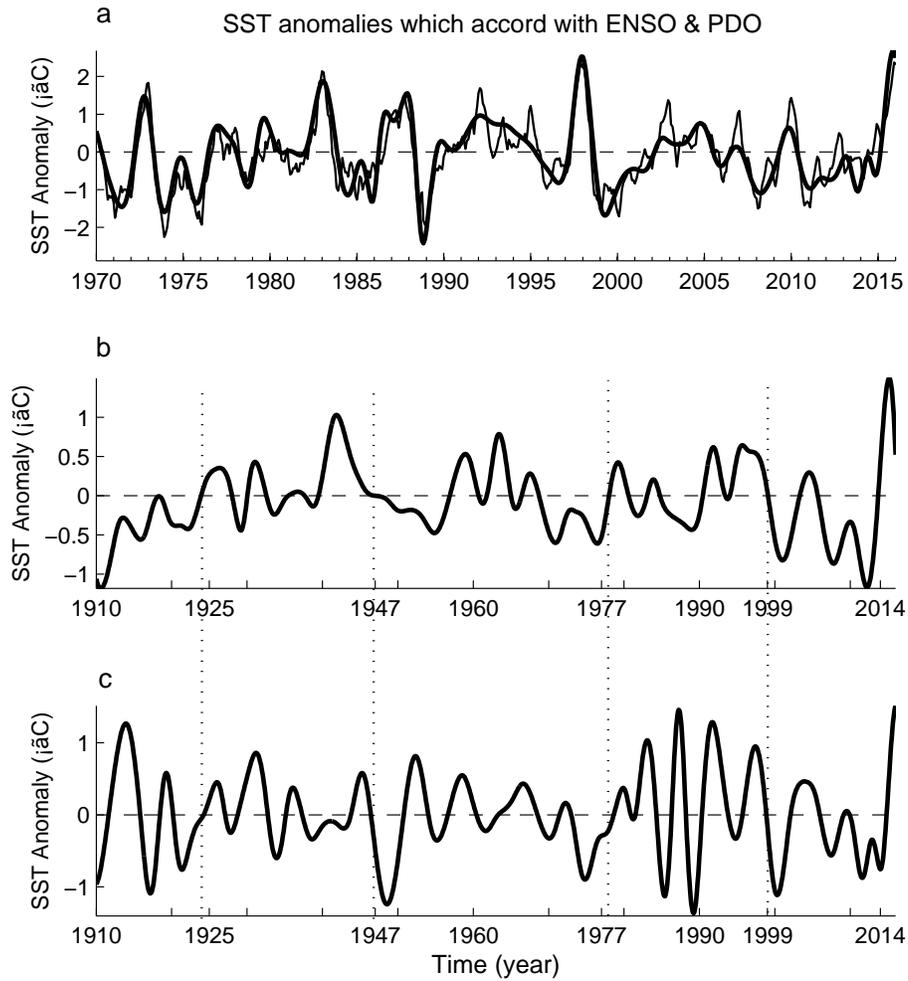}}

\caption{Feasibility of the decompositions by ESMD method. \textsf{a}: the comparison between the sum of Mode 2-5
for point A (thick curve) and the known Ni\~{n}o-3.4 index (the average SST anomaly in the region bounded by $5^{\circ}$N- $5^{\circ}$S from $170^{\circ}$W to $120^{\circ}$W, denoted by a thin curve); \textsf{b} and $\textsf{c}$ display the sum of Mode 3-6 for point B and that of Mode 3-5 for point A separately. The four vertical dashed lines correspond to the selected zero-points in \textsf{b}.}
\end{figure}

\begin{figure}[h]
 \centerline{\includegraphics[width=\textwidth]{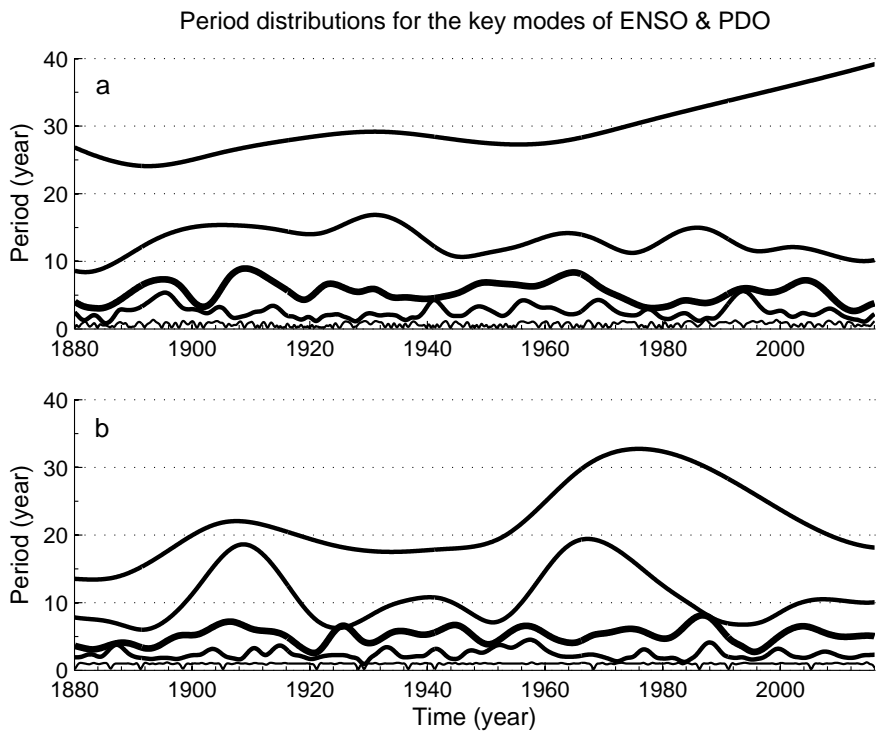}}

  \caption{Period distributions of the modes along the time. \textsf{a} and \textsf{b} are associated with points A and B separately. In each subgraph the $5$ curves from below to above correspond to Mode 1-5 accordingly (that for Mode 6 to B is omitted here). The thickest curves here denote the variations of mutual key mode for ENSO and PDO with half-decadal time scale.}
\end{figure}

\begin{figure}[h]
 \centerline{\includegraphics[width=\textwidth]{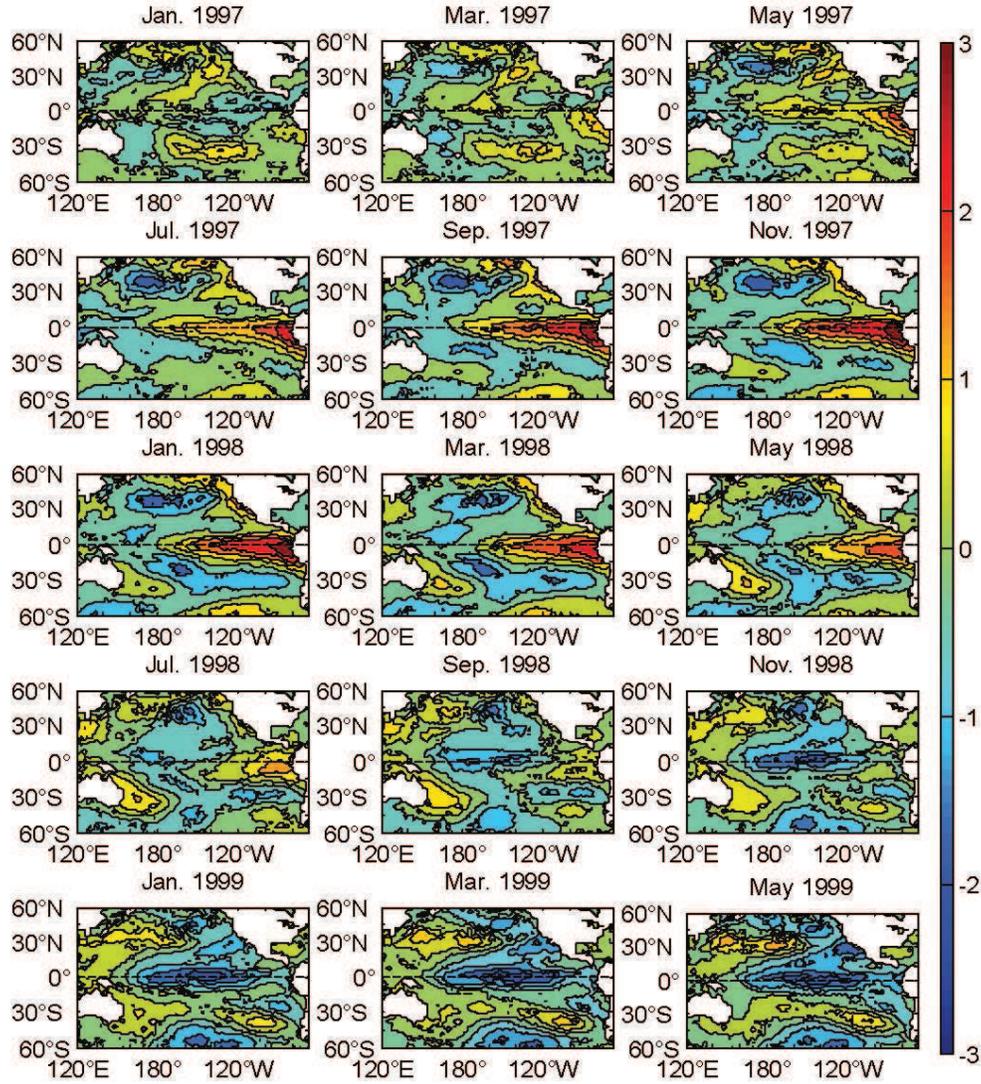}}

  \caption{A series of spatial patterns given by the integrated ESMD method which reflect the evolution process of ENSO. Here
only the seasonal and global-warming signals are abandoned. The color bar has a unit $^{\circ}$C.}
\end{figure}

\begin{figure}[h]
 \centerline{\includegraphics[width=\textwidth]{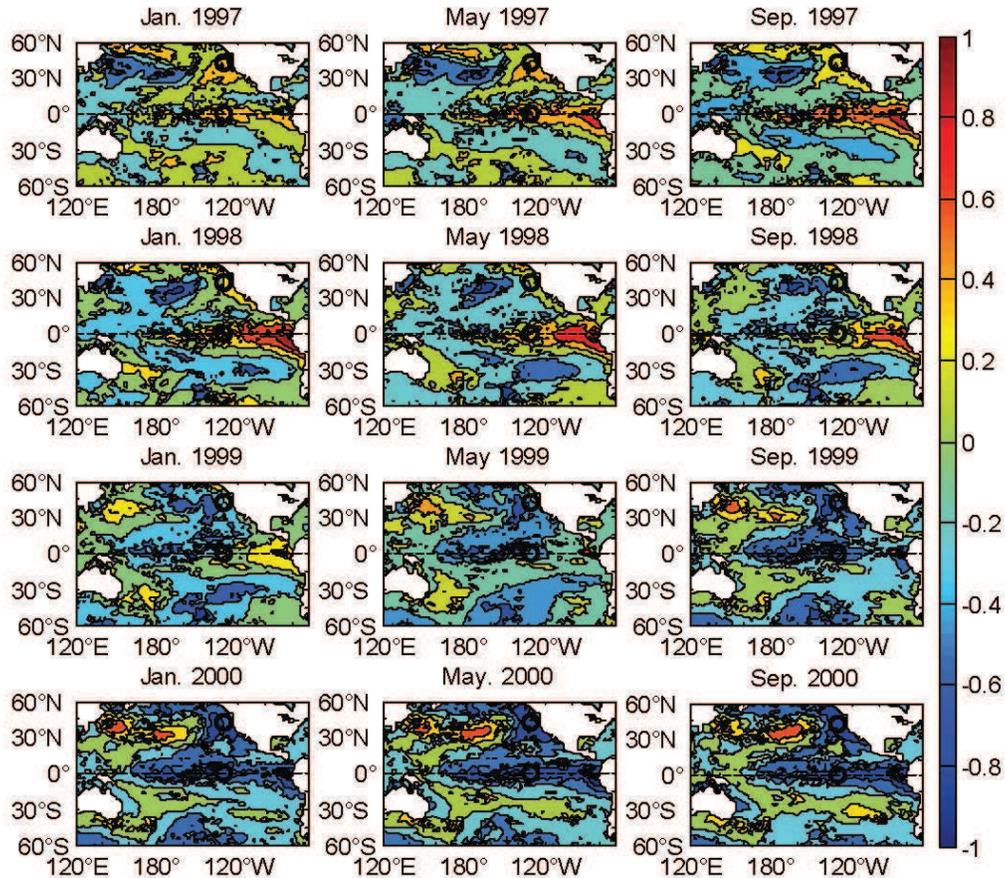}}

  \caption{A series of spatial patterns given by the integrated ESMD method which reflect the evolution process of PDO's regime-shift. Here only the half-decadal mode (Mode 3) is used. The locations of the lower and upper circles correspond
to grid points A ($0^{\circ}$, $130^{\circ}$W) and B ($42^{\circ}$N, $130^{\circ}$W) separately. The color bar has a unit $^{\circ}$C.}
\end{figure}

\end{document}